# Can Inorganic Salts Tune Electronic Properties of Graphene?

Guilherme Colherinhas,[1] Eudes Eterno Fileti[2] and Vitaly V. Chaban[2]

1) Departamento de Física, CEPAE, Universidade Federal de Goiás, CP.131, 74001-970, Goiânia, GO, Brazil

2) Instituto de Ciência e Tecnologia, Universidade Federal de São Paulo, 12247-014, São José dos Campos, SP, Brazil

**Abstract**. Electronic properties of graphene quantum dots (GQDs) constitute a subject of intense scientific interest. Being smaller than 20 nm, GQDs contain confined excitons in all dimensions simultaneously. GQDs feature a non-zero band gap and luminesce on excitation. Tuning their electronic structure is an attractive goal with a technological promise. In this work, we apply density functional theory to study an effect of neutral ionic clusters adsorbed on GQD surface. We conclude that both HOMO and LUMO of GQD are very sensitive to the presence of ions and to their distance from the GQD surface. However, the alteration of the band gap itself is modest, as opposed to the case of free ions (recent reports). Our work fosters progress in modulating electronic properties of nanoscale carbonaceous materials.

**Key words**: graphene, ions, ion pair, band gap, electrical properties, density functional theory.

TOC IMAGE

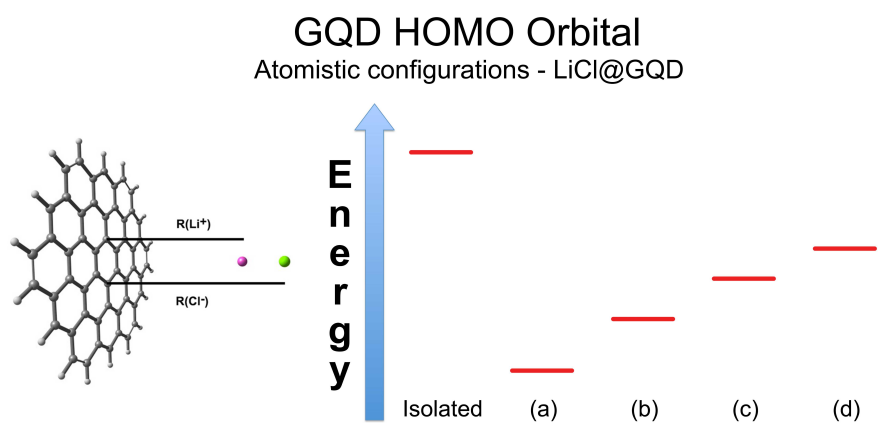

**Introduction**

Graphene is currently one of the most researched nanoscale entities because of its interesting electrical and optical properties and two-dimensional structure.[1-4] Graphene sheets can be doped, functionalized and self-organized offering significant technological promise.[2, 5-14] Being a zero-overlap semi-metal (both electrons and holes are charge carriers), graphene generates research interest in the context of various applications, where high conductivity is a bottleneck. Four electrons of carbon are able to participate in chemical bonding, whereas only three of them appear actually employed. The remaining electron (π-electron) is available for electricity conduction. Experiments showed that the electronic mobility of graphene is notably high (ca. 15,000 $cm^2 \cdot V^{-1} \cdot s^{-1}$).[15, 16] In turn, the theoretical limit amounts to 200,000 $cm^2 \cdot V^{-1} \cdot s^{-1}$.[17] The listed value was computed from scattering of acoustic photons. Electrons of graphene resemble photons in many aspects. For instance, they do not scatter while moving over sub-micrometer distances.[18] It is believed that the major factor limiting conductivity of graphene is the quality of its structure. If this supposition is true, conductivity can be essentially improved in the future.

Most electronic applications of graphene are handicapped by the absence of band gap. The band gap of graphene can be tuned using dopants, which is probably the most feasible method nowadays permitting to modulate semiconducting properties of materials.[11, 19-21] Boron and nitrogen atoms are often employed for this purpose.[20, 22] The substitutional doping is another popular and powerful approach to enlarge the band gap. The doped graphene exhibits electronic structure with a linear dispersion relation similar to pristine graphene.[23] However, band gap can be opened through substitutional doping using boron and nitrogen. The Fermi level appears within valence band when boron is employed and within conduction band when nitrogen is employed. Consequently, p-type and n-type semiconducting electronic properties can be achieved. This feature is important for application in electronic devices.

Hole acceptor or electron donor can be attached to graphene modifying its electronic structure. This sort of manipulation is known as chemical modification. For instance, if graphene is modified by nitrogen dioxide or ammonia, they will generate charge carriers. Similarly, adsorbed water molecules play a role of defects. The defects, in turn, help tuning the band gap.

Graphene quantum dots (GQDs) are nothing else than small fragments of graphene, where electron transport is confined (due to size of fragment) in all spatial dimensions. Their band gap is tunable through modification of size and surface chemistry.[5, 13, 20, 24-27] Whereas the band gap of benzene is about 7 eV,[28] GQDs feature significantly smaller values (see below). Major limitations of GQDs include poor dispersion in the organic and inorganic solvents and aggregation.[20, 24, 25]

This work investigates a non-covalent modification of GQDs. We use inorganic neutral clusters of ions (ion pairs) and quantitatively describe their effect on the electronic structure of GQD. Recently, Kalugin and coworkers[29] demonstrated that the energy of the 2s molecular orbital of lithium falls between the highest occupied molecular orbital (HOMO) and lowest unoccupied molecular orbital (LUMO) of the semiconducting carbon nanotube. It was suggested that lithium ions (employed in the electrochemical devices) can be used to efficiently tune the semiconducting band gap of the nanotube. It remained, however, unclear whether a charged periodic system constitutes a perfect computation model for the phenomenon in question. Kalugin and coworkers used background charge to neutralize their system, which may affect electronic structure in sometimes unpredictable manner. Valence and conduction bands of semiconductors are extremely sensitive to computational methodology. Furthermore, only a few methods are able to reproduce the experimental band gap numerically.

Colherinhas and coworkers[30] systemized the effect of light ions. They avoided charged periodic systems and demonstrated that 2s (Li) and 3s (Na) molecular orbitals become LUMOs for the semiconducting graphene sheet. Therefore, the band gap can be decreased depending on the distance of the ion to the plane of graphene-based structure. This work also did not account for the counter-ion and solvent molecules, which influence the electronic structure independently.

Our current results suggest that the anion (chloride) induces principal changes, as compared to the charged (periodic and non-periodic) systems. The polarizing effect of $Li^+$, $Na^+$, $Mg^{2+}$ in relation to GQD decreases. LUMOs of ions obtain higher energy, which in most cases appears above LUMO of GQD. We conclude that anions must be removed (that is, ion pairs destroyed) to tune electronic properties of GQDs, and probably other graphene-based derivatives. Charged electrodes provide the coveted working conditions by charge separation in the external electric field.

**Methodology**

This work reports a series of single-point pure density functional theory (DFT) calculations involving the model graphene sheet, the lithium, sodium, magnesium cations and the chloride counter-ion (Figure 1). The graphene sheet was terminated by hydrogen atoms. We deliberately avoid using an infinite model of graphene, since sufficiently large graphene sheets are zero-gap semiconductors. The wave functions were constructed using the LANL2DZ[31] basis set within the framework of the BLYP functional (generalized gradient approximation).[32] It is known that pure DFT methods tend to overestimate an electron transfer. In turn, modest basis sets demonstrate an opposite trend. Based on our preliminary tests, certain compensation occurs. One can expect that the results reported here are in concordance with those obtained using hybrid DFT and a comprehensive basis set. Furthermore, significantly large systems, such as those investigated in the present work, require relatively modest level of theory for efficient electronic-structure computations. Apart from technical considerations, self-consistent field (SCF) convergence using comprehensive basis sets can be very problematic in the case of graphene due to its specific conduction band. Simulated configurations are summarized in Table 1.

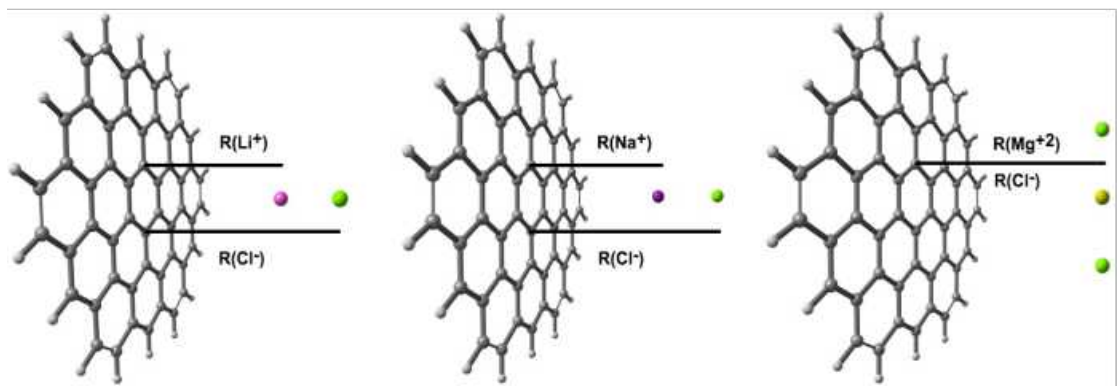

Figure 1. Simulated molecular configurations. The cations ($Li^+$, $Na^+$, $Mg^{2+}$) were placed at distances 0.25, 0.45, 0.65 and 0.85 nm from GQD. The distances between cation and anion in LiCl, NaCl and $MgCl_2$ were estimated based on the geometry optimization at the BLYP/LANL2DZ level of theory.

The energy-based criterion of the SCF convergence was set to $10^{-8}$ Hartree in all systems. If convergence was not achieved after 128 cycles, the quadratically convergent SCF procedure was called. The method was introduced in Ref. [33, 34] This method involves linear searches when far from convergence and Newton-Raphson steps when close to convergence. The quadratically convergent SCF procedure performs much slower than regular SCF, but eventually converges in many difficult convergence cases. Basis set superposition error was estimated and deducted from the binding energy using the counterpoise approach. All electronic structure calculations were performed in Gaussian 09 (*www.gaussian.com*).[35]

Table 1. Simulated atomistic configurations. $R(Li^+/Na^+/Mg^{2+})$ and $R(Cl^-)$ indicate distances (in nm) of the corresponding particles from GQD. Four configurations – (a), (b), (c), (d) – were considered for each of the three chemical compositions: LiCl@GQD, NaCl@GQD and $MgCl_2$@GQD

| Setup | $R(Li^+)$ | $R(Cl^-)$ | $R(Na^+)$ | $R(Cl^-)$ | $R(Mg^{2+})$ | $R(Cl^-)$ |
|---|---|---|---|---|---|---|
| (a) | 0.25 | 0.45 | 0.25 | 0.49 | 0.25 | 0.25 |
| (b) | 0.45 | 0.65 | 0.45 | 0.69 | 0.45 | 0.45 |
| (c) | 0.65 | 0.85 | 0.65 | 0.89 | 0.65 | 0.65 |
| (d) | 0.85 | 1.05 | 0.85 | 1.09 | 0.85 | 0.85 |

(isolated) - reference graphene sheet terminated by hydrogen atoms to satisfy valence

**Results and Discussion**

Figure 2 provides analysis of energy levels of HOMOs and LUMOs of pristine GQD and upon adsorption of inorganic ions (neutral clusters). The energies of orbitals are given as a function of ionic positions in relation to GQD (Table 1). HOMO (-4.46 eV) and LUMO (-2.56 eV) of pristine GQD are provided for easier detection of changes due to coordination of the cations and the anion.

The distance between GQD and cation was increased gradually from 0.25 to 0.85 nm (Table 1). The distance between the cation and the anion was optimized to correspond to minimum internal energy, according to the applied DFT method and basis set. Therefore, the listed distances of the chloride anion to GQD are different in its clusters with $Li^+$, $Na^+$ and $Mg^{2+}$. In the case of $MgCl_2$, the anions were placed on both sides of the cation. These modeling setups allow investigating how (1) the ion pair influences electronic density distribution on graphene; (2) how the effect, if any, depends on distances between GQD, cation and anion. All ion pairs decrease energy of both valence and conduction bands of GQD. This trend means that binding of small ionic clusters to GQD is definitely energetically favorable. The effect is strongly dependent on the nature of cation. Compare, LiCl decreases HOMO of GQD by 0.33 eV, NaCl by 0.41 eV and $MgCl_2$ by 0.06 eV. These values correspond to setup (a). As ionic cluster moves farther away, the effect decreases smoothly. In the case of LiCl, energy gain of HOMO due to complexation with inorganic ions constitutes 0.33, 0.25, 0.19, 0.14 eV for separation distances of 0.25, 0.45, 0.65, 0.85 nm. The observed changes are significant influencing also total potential energy of each system.

Lithium ion must be considered more polarizable than sodium ion. Nevertheless its effect on HOMO and LUMO of GQD (Tables S1-S2) appears smaller. We hypothesize that lithium engenders a strong bond with chlorine, involving certain degree of covalence. As a result of this bonding, valence electronic density on the cation is adjusted. The polarizing action (electric field) is decreased. In turn, binding of $Na^+$ with $Cl^-$ is weaker, whereas influence of sodium cation of GQD is stronger.[36] This effect is, to a significant degree, unexpected constituting a poorly predictable interplay of many-body interactions in the ion-molecular system.

The case of $Mg^{2+}$ is significantly different from the case of light alkali ions. First, presence of $MgCl_2$ changes HOMO energy of GQD by just 0.06 eV, being much smaller than the alkali ions do. Second, although HOMO and LUMO decrease when $Mg^{2+}$ is close, setup (a), larger separations result in very similar energy alterations. Therefore, $Mg^{2+}$ is not a robust choice to tune electronic structure of GQD. The same trend would be probably observed in all other nanoscale carbonaceous structures with a finite band gap.

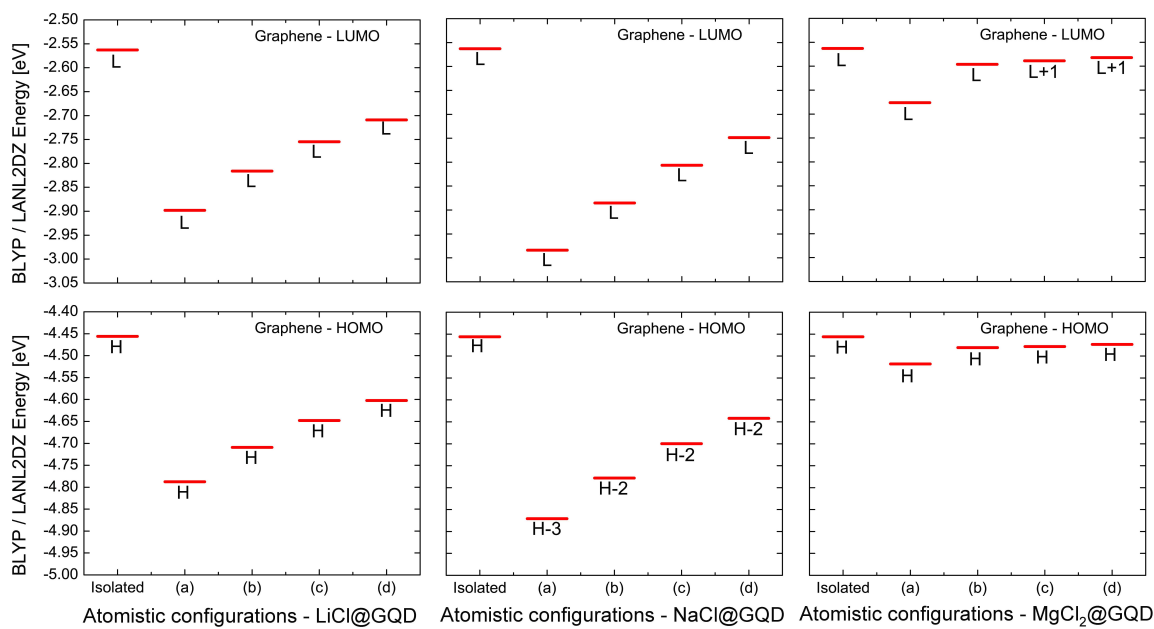

Figure 2. HOMO and LUMO energy levels of GQD in the LiCl@GQD, NaCl@GQD, and MgCl$_2$@GQD in the investigated setups (see Table 1). Note that the considered orbitals are those localized on GQD.

Unlike one recently showed for lonely cations,[30] LUMO of LiCl and NaCl appear above LUMO of GQD (Tables S1-S2). LUMO of MgCl$_2$ is also above LUMO of GQD (Tables S3) when the ion cluster is near GQD, but drops below when Mg$^{2+}$ is moved farther (0.65, 0.85 nm from GQD). Adsorption of ions increases their LUMOs systematically. LUMO of the isolated LiCl exhibits energy of -2.18 eV, whereas it is increased up to -0.18 and -1.90 eV in the setups (a) and (b) respectively. Similarly, LUMO of the isolated NaCl is at -1.91 eV, whereas it at -0.09 and -1.26 eV in the setups (a) and (b). In the case of MgCl$_2$, the effect is much smaller, -2.51 eV (isolated MgCl$_2$) vs. -1.84 eV, setup (a), and -2.52 eV, setup (b). Increase in atomic mass and electron charge of the cation results in lower LUMO energy. Accordingly, LUMO of cation can fall between HOMO and LUMO of GQD provided that a heavier alkali metal is chosen, e.g. rubidium or cesium. The energy of LUMO in any ion pair is greatly sensitive to its distance to GQD. This observation suggests a strong binding between the considered species.

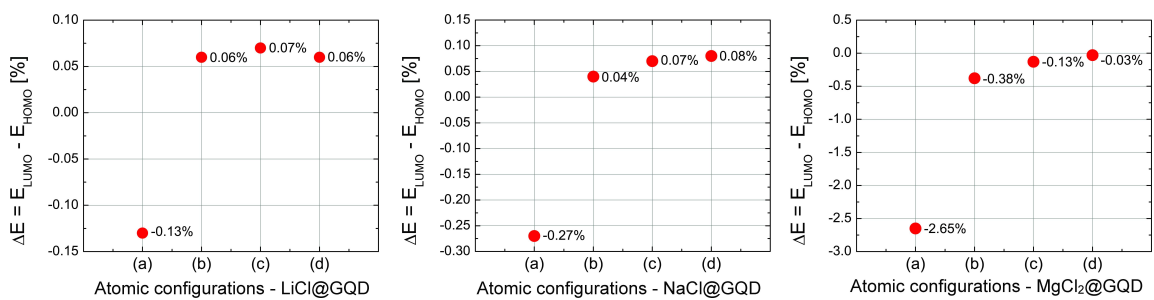

Figure 3. Alterations of the HOMO-LUMO band gap of GQD in the presence of LiCl, NaCl and MgCl$_2$.

Visualized HOMO and LUMO in the LiCl@GQD (Figures 4), NaCl@GQD (Figure 5) and MgCl$_2$@GQD (Figure 6) systems indicate localization of molecular orbitals in GQD and ion pairs as a result of their binding. LUMO of the system is primarily localized on GQD. Molecular orbitals of ions lie either well below or above delocalized p-electrons (fourth electron of each carbon) of the nanoscale carbon. Figure 5 indicates that only LUMO+15, setup (a), and LUMO+2, setup (d), are localized on ions, whereas all lower-energy unoccupied molecular orbitals are observed on GQD. Similar trend was observed in NaCl@GQD, where LUMO of ions is well above LUMO of GQD, nevertheless strongly depending on their distance to GQD.

The case of MgCl$_2$@GQD is somewhat different. LUMO+4 in the setup (a) is shared by ions and GQD, while LUMO in the setup (d) is localized exclusively on MgCl$_2$. Accordingly, tuning MgCl$_2$—GQD separation represents a very interesting opportunity to modulate electronic properties of GQD using the same working salt. Similarity of LUMO in GQD and MgCl$_2$ is largely due to a strong interaction between these chemical entities, which happens, in turn, due to a strong polarization by a divalent cation.

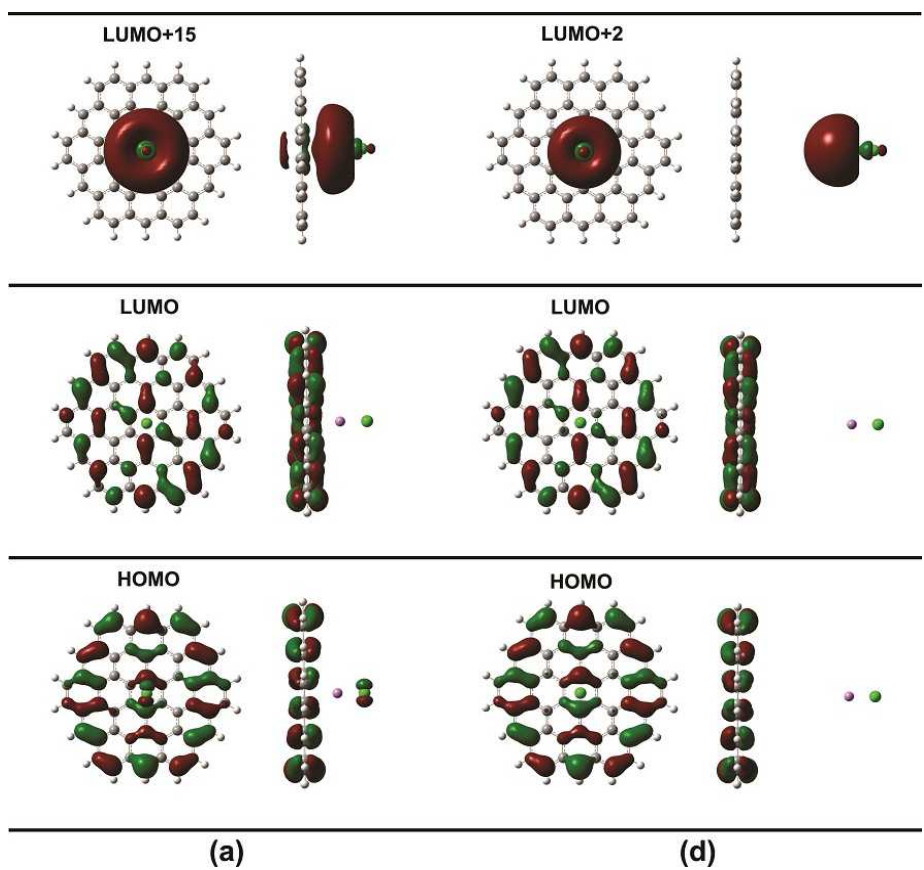

Figure 5. HOMO and LUMO of LiCl@GQD in setups (a) and (d). Setups (b) and (c) are similar to (d).

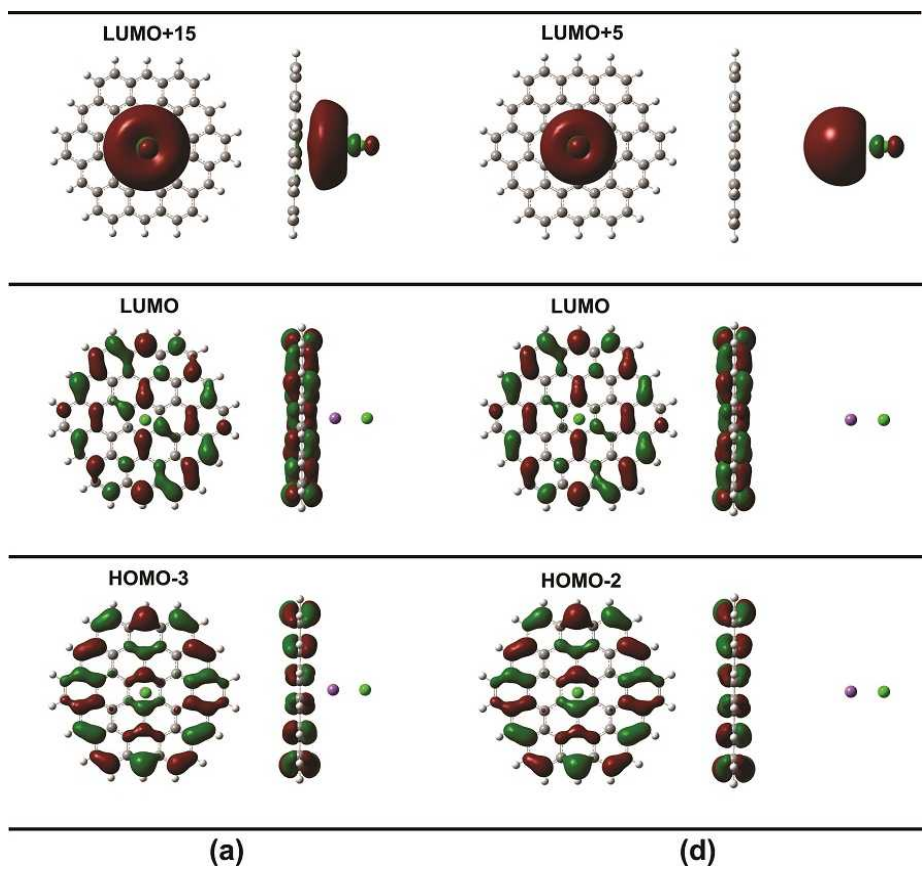

Figure 6. HOMO and LUMO of NaCl@GQD in setups (a) and (d). Setups (b) and (c) are similar to (d).

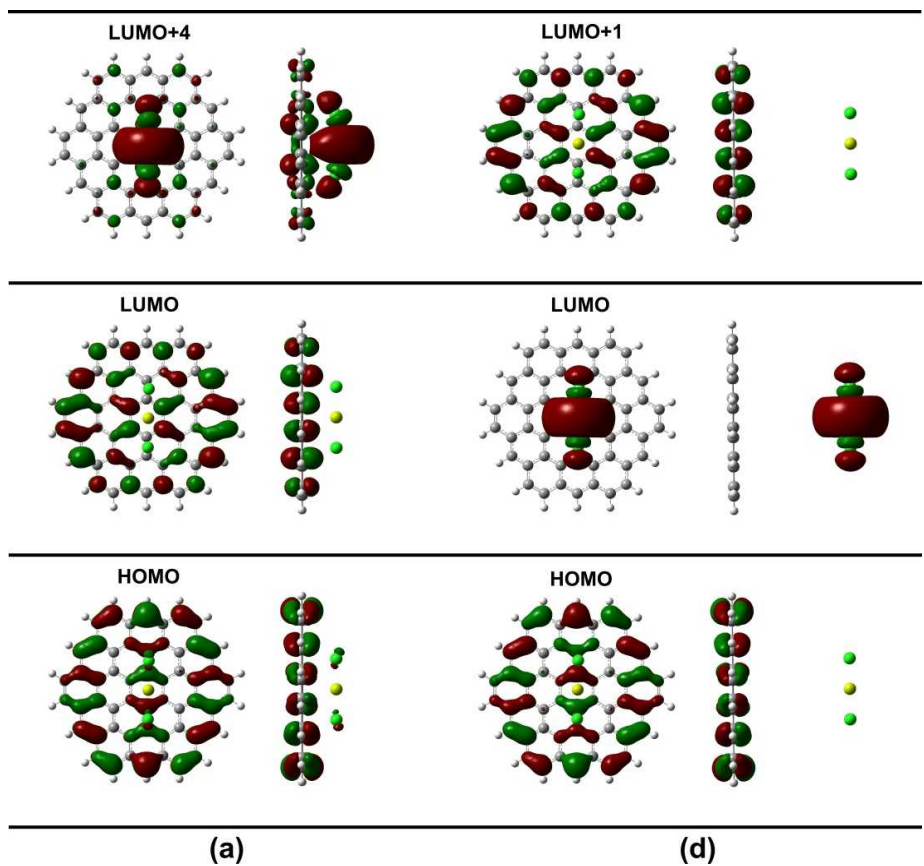

(a)    (d)

Figure 7. HOMO and LUMO of MgCl$_2$@GQD for setups (a) and (d). Setups (b) and (c) are similar to (d) and are omitted for this reason.

Our hypothesis about strong binding of small ionic cluster to GQD is completely confirmed by the analysis of binding energies (Figure 8). The absolute values of interaction energy suggest that interaction goes beyond van der Waals (VDW) term. Purely VDW binding would result in a much weaker attraction. Decay of energy, *E(r)*, with respect to distance, *r*, was reliably correlated to the Coulombic attraction, $E(r) \sim r^{-2}$. According to non-linear regression, correlation coefficient equals to 0.99. The inorganic ions induce partial electrostatic charges on the surface of GQD, which are responsible for the Coulombic behavior of the energy curves (Figure 8).

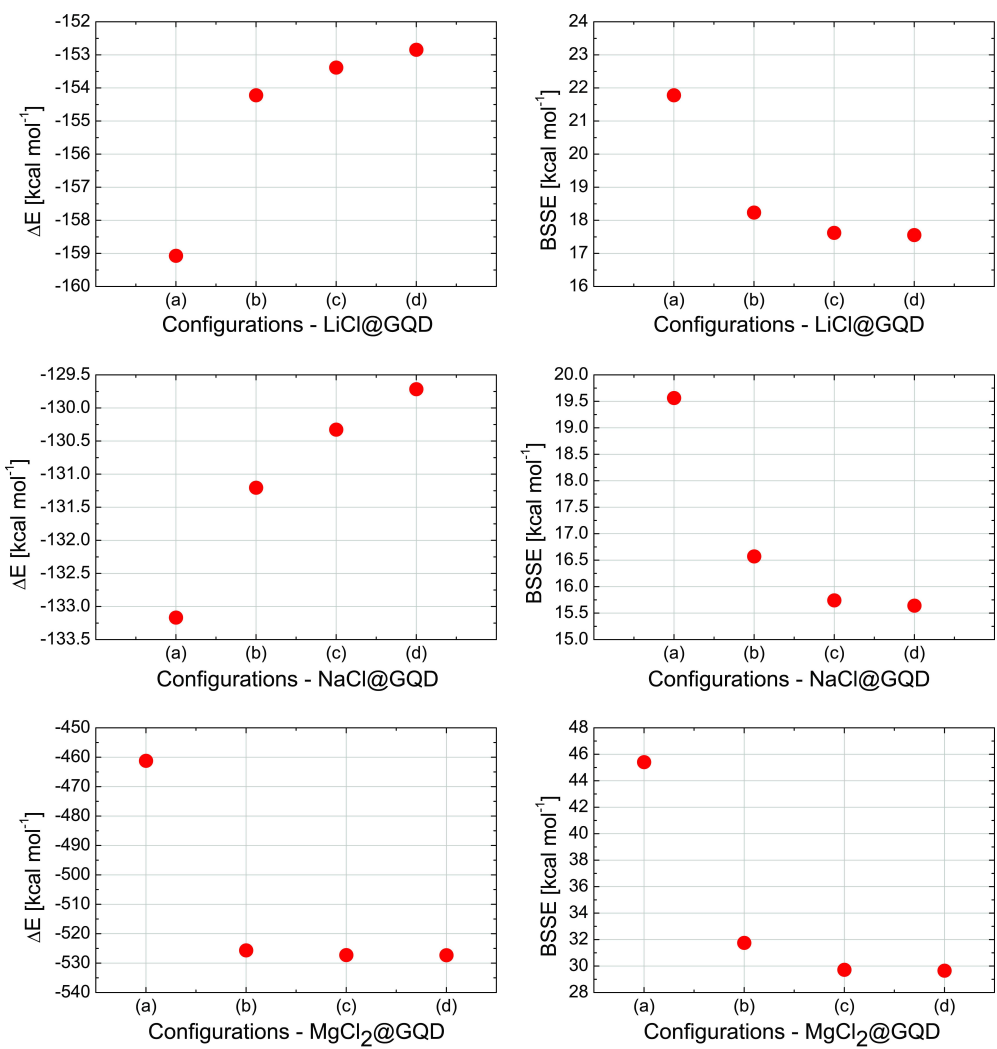

Figure 8. Binding energy keeping GQD and neutral ion cluster together (left) and basis set superposition error (BSSE, right) in all the investigated setups.

The basis set superposition error (Figure 8) is proportional to binding energy constituting 10 to 15% of the latter. BSSE is not only a by-product of the calculation, but also an interesting measure of how much one chemical species benefits from the basis set of another chemical species. For instance, carbon atoms of GQD are able to employ 3s orbitals of sodium and 3s and 3p orbitals of chlorine. This must increase superposition of functions in NaCl@GQD, as compared to other systems. Indeed, BSSE is detected to be largest in the NaCl@GQD system. BSSE can be generally decreased if the total number of relevant functions in the selected basis set is increased.

The introduced results have been obtained using pure density functional theory and relatively modest basis set with soft-core potentials to adequately represent non-valence electrons. This was done due to a significant number of electrons in the considered ion-molecular systems. In terms of electron transfer, the flaw of pure density functional (favors electron delocalization) is compensated by modest basis set (underestimates electron delocalization). Nevertheless, comparison to a more accurate method may be sometimes important. We provide HOMO as a function of separation in Figure S3 derived using hybrid density functional theory and the Pople-type basis set including polarization functions, 6-31G(d). As expected, a larger basis set resulted in lower energies of all orbitals (HOMO energies are depicted in Supplementary Information), nevertheless the trend remained unaltered. We believe that our methodological choice is well-justified.

It must be noted that the band gap of GQD is strongly dependent on its finite size and terminating atoms. Unlike graphene, which exemplifies a zero-gap semiconductor due to its macroscopic length in the two dimensions, GQDs features a large HOMO-LUMO band gap, which increases as the size of GQD decreases.

**Conclusions**

We reported density functional theory investigation of the ion-pair@GQD complexes. We showed that clusterized ions are able to significantly polarize electron energy levels of GQD. Furthermore, the polarizing action is a function of (1) ion position and (2) counter-ion position. If solvent is absent, the polarizing action persists at significant separations, 8.5 nm from GQD.

Our results constitute interest and importance in the context of electrochemical applications of GQDs and its derivatives and sensor setups. Electronic structure is important for chemical reactivity and electrical conductivity. An ability of alkali metal ions to position their LUMOs between HOMO and LUMO of pristine graphene is not trivial. It allows either intelligent tuning of electrical properties of these systems or predicting alterations in the physical chemical properties in the manifold ionic environments. However, the cation and the anion must be spatially separated, as the present results indicate.

**Supplementary Information**

Tables S1-S4 and Figure S1 summarize energies of molecular orbitals in the ion cluster@GQD systems. Table S5 reports binding energies. Figure S2 visualized selected molecular orbitals. Figure S3 provides simulation results using higher level of theory. This information is available free of charge via the internet.

**Acknowledgments**

G.C. and E.E.F. thank Brazilian agencies FAPESP and CNPq for support. V.V.C. acknowledges research grant from CAPES under the "Science Without Borders" program.

**AUTHOR INFORMATION**

E-mail addresses for correspondence: gcolherinhas@gmail.com (G.C); fileti@gmail.com (E.E.F.), vvchaban@gmail.com (V.V.C.)